\documentclass{article}
\usepackage{epsfig}
\usepackage{caption}
\usepackage{latexsym}
\usepackage{amsmath}
\newcommand{\de}{\partial}
\renewcommand{\a}{\alpha}
\renewcommand{\b}{\beta}
\newcommand{\beq}{\begin{equation}}
\newcommand{\eeq}{\end{equation}}
\newcommand{\bea}{\begin{eqnarray}}
\newcommand{\eea}{\end{eqnarray}}
\begin{document}



\title{Holography and Emergent 4D Gravity}

\author{Francesco Nitti\\
Centre de Physique Th\'eorique, Ecole Polytechnique,\\
91128, Palaiseau, France}

\maketitle


\begin{abstract}
\noindent
I review recent work  toward constructing, via
five-dimensional  holographic duals,  four-dimensional theories in which
spin-2 states (gravitons) are emergent. 
The basic idea is to  extend to gravity model-building the
applications  of holographic  duality to  phenomenology construction.
\end{abstract}


\section{Introduction}


This article  is a brief review of the work Ref. \cite{kn},
in which we applied  the ideas of $AdS$/CFT to model-building
of theories with modified gravity in four dimensions. 

The $AdS$/CFT correspondence \cite{maldacena,witten,gkp} establishes an
equivalence between  a higher-dimensional, gravitational theory on a curved
background,  and a lower-dimensional field theory,  without  dynamical gravity.

The basic question I will address here is,
{\em how can one obtain dynamical gravity also on
 the field theory side,} i.e. in the  {\em lower dimensional} theory?
More specifically, is it possible
to use the  bulk theory  as a model for observed gravity ? 

The interest in  this question arises because one could hope to use
 holography to construct a UV-complete theory
of gravity, as well as consistent modified gravity
theories in four-dimensions, e.g. models  which are
 UV-complete and  contain  massive or  metastable gravitons. Models with such properties
are difficult to construct directly in four-dimensions, and often
suffer from inconsistencies. These difficulties could be bypassed by defining
 models with the desired  properties via holographic duals, which
propagate  {\em ordinary} Einstein gravity in higher dimensions.

A simple case of a five/four duality
in which the 4D  theory contains
dynamical gravity is the Randall-Sundrum (RS) model \cite{RS1,RS2}
 (see Refs. \cite{maldatalk,gubser,apr,zaffa} for its holographic
interpretation).
 The 5D
theory is defined in  a cut-off  Anti-de Sitter ($AdS$)
 space-time. 
The role of the 4D graviton is played by
a normalizable zero-mode, whose   profile in the radial direction
of $AdS_5$ is peaked in the UV region,
as shown in figure 1 (a). This signals the fact that
the 4D graviton in RS is a fundamental degree of freedom.

The holographic dual of
the  RS model, therefore,  does have
 four-dimensional dynamical gravity, but this
is achieved trivially by adding the graviton as a fundamental
degree of freedom. In particular, in this model
the high energy behavior of 4D gravity is the standard one.

It would be much more interesting to have a situation like
the one sketched in figure 1 (b), in which the graviton
zero-mode profile is peaked in the IR region: in this case, by  standard
holography arguments, the four-dimensional graviton
would be seen as a composite, rather than a fundamental degree
of freedom, and gravity would be an emergent manifestation
 of some strong, {\em non-gravitational}
IR dynamics. In such a situation, the graviton will cease
to exist at high energy, making gravity soft in the UV. This
would be a concrete realizations of  the composite graviton ideas of 
Ref. \cite{elias1,elias2}, and the ``fat graviton'' ideas
of  Ref. \cite{sundrum}


\begin{figure}[h]
\begin{center}
\leavevmode \epsfxsize=6cm \epsffile{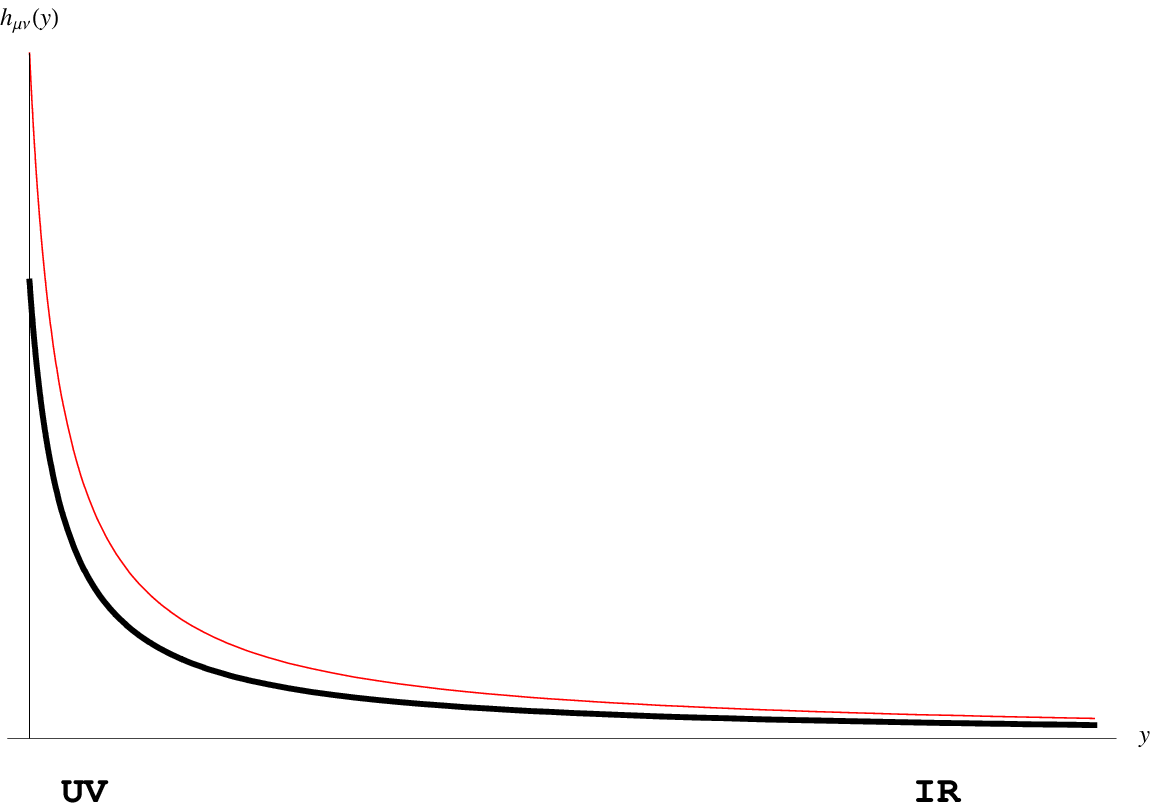}\hspace{0.5cm}
\leavevmode \epsfxsize=6cm \epsffile{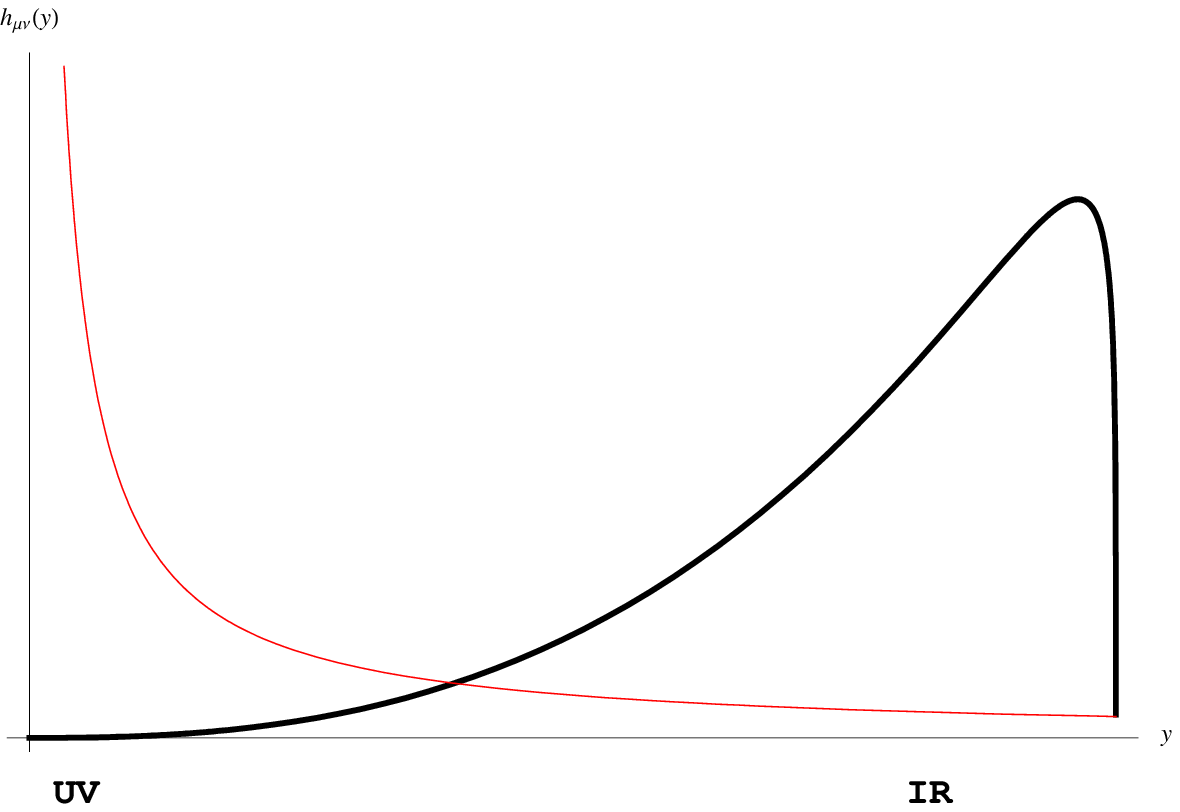}\\
(a)\hspace{6cm}(b)
\end{center}
\caption{A sketch of the graviton profile (thick black curve)
in  two possible holographic realizations
of four-dimensional gravity:
(a) the graviton zero-mode profile in RS2, as a function
of the AdS radial coordinate $y$; (b) how the same profile
would look like if the graviton were a composite state. In
both figures the thin red curve is a sketch of the scale factor of the
5D background. The UV and IR correspond to the large and small
scale factor regions, respectively.\label{fig1}}
\end{figure}

  The problem of obtaining  emergent gravity in 4D via holography
was addressed in a very general class of five-dimensional
models   in Ref. \cite{kn}. There,  we assumed a bottom-up  approach,
and used the ``phenomenological''
version of the holographic setup in which, rather than a
fully string-theoretical construction, the holographic model
consists of Einstein gravity (plus eventually other fields) in a
 warped five-dimensional background. This approach
is not as rigorous as the ``stringy'' $AdS$/CFT correspondence,
nevertheless it is believed to be a valid tool in model building
 (see Ref. \cite{ghergrev} for a recent review).

This work is organized as follows. In Section 2, I review the ideas
that underlie  the holographic approach to phenomenology. In Section
3, I discuss 4D graviton states in RS-like models. Sections 4 to 8
are devoted to the main results obtained in Ref. \cite{kn}:
 in Sections 4 and 5 I discuss  the possibility
of obtaining  emergent massless 4D spin-2 modes  from  generic asymptotically
$AdS_5$ space-times; Section 6 discusses lower spin states; in  Section 7
I briefly discuss how to include matter in the setup; a critical
analysis  of the results and  possible
generalizations  are discussed in Section 8.

In this article I will omit  most  of the calculations.
The reader is referred to Ref. \cite{kn} for all the details,
and for a more complete list or references.

\section{Holography and Phenomenology}
\setcounter{equation}{0}

After the original $AdS$/CFT conjecture, it was realized that holography can be effectively
used for model building, and people  started thinking  about extra-dimensional models
in terms of holographic duality \cite{maldatalk,gubser,apr,zaffa}

The starting point for ``holographic phenomenology'' is the following.
According to gauge/gravity duality, the particle content on the field
theory side can be read off
as the spectrum of normalizable\footnote{In a sense to be made more precise
below.}
fluctuation modes  around the gravity dual background \cite{witten}.
We will take the latter
to be a 4+1-dimensional space-time, labeled by coordinates ($x^\mu, y$).
Then, a
normal mode $\phi_{\{\alpha\}}(x^\mu,y)$ (where $\{\alpha\}$ denotes
a set of quantum  numbers), such that
\beq
\Box_4 \phi_{\{\alpha\}}(x^\mu,y) = m^2 \phi_{\{\alpha\}}(x^\mu,y),
\eeq
corresponds to a state of mass $m$ in the dual gauge theory, with the same quantum
numbers $\{\alpha\}$.

The normalizability requirement for a given  mode
 means essentially that one can  ``dimensionally reduce''
the action for that mode by integrating over $y$, and obtain a finite 4D
effective kinetic term.
 This condition  corresponds to normalizability of the wave function in
an appropriate quantum mechanical problem.

In the holographic approach to phenomenology, one can
therefore ``engineer'' the 5D space-time in such a way as to
obtain a 4D spectrum with desired features. In this
language,  the five-dimensional description of the model is more tractable than
the four-dimensional one, which is that of  a (generically
unknown) strongly coupled gauge theory.
The 5D fluctuation equations  compute directly the spectrum of gauge-invariant 
states,
and effectively the 4D model is defined though its 5D dual (see Ref. \cite{ghergrev}
for a recent review and list of references).

The simplest example is given by the Randall-Sundrum (RS) models \cite{RS1,RS2}, in
which gravity and eventually  other fields live  in a slice of $AdS_5$ space,
\beq\label{ads}
ds^2 = {1\over (ky)^2} \left(dy^2 + \eta_{\mu\nu}dx^\mu dx^\nu\right),
\qquad y_{UV} < y < y_{IR}.
\eeq
The coordinate $y$ is dual to the inverse energy scale in 4D, $E\sim y^{-1}$; the small  and
large  $y$ regions correspond to the UV and the IR, respectively
of the dual 4D theory; the boundaries  $y_{UV}$ and $y_{IR}$ correspond to UV and IR cut-offs,
 and break conformal invariance;
different choices of  boundary conditions for the various fields yield  different 4D spectra.

One important  question, when one considers a particular fluctuation mode,
regards  its localization
properties: modes whose profile is peaked  in the small $y$ region are dual to
elementary states in the 4D picture; on the other hand, modes that are peaked in the IR
correspond to composite states. This distinction is used in model building, for example,  to
construct models in which various SM particles have different compositeness properties, by
localizing  them in different regions of the $AdS$ slice.

In an asymptotically $AdS_5$ space (i.e. when the UV cut-off is removed, $y_{UV}=0$,
and the theory is UV-complete), normalizable
states typically vanish in the UV, and are therefore composite. In general, any mode
has an expansion, near the $AdS$ boundary $y=0$, of the form:
\beq
\phi(x,y) \sim y^{\Delta_-} \phi_-(x) +  y^{\Delta_+} \phi_+(x) +\ldots, \qquad y\to 0,
\eeq
where $\Delta_- < \Delta_+$ depend on the equation governing the particular fluctuation.
Typically $\phi_-$-modes are not normalizable, and correspond to external sources in the
 4D dual FT, while  $\Phi_+$-modes are normalizable (around $y=0$) and correspond to IR modifications of the theory.

Here, we are  interested in 4D states that have the same quantum numbers
of physical gravitons in 4D, i.e. transverse and  traceless symmetric tensors
under the 4D Lorentz group. These can be naturally identified, in the gravity dual,
with the tensor fluctuations $h_{\mu\nu}(x,y)$ of
the  components along space-time of the 5D metric, satisfying:
\beq
\de^\mu h_{\mu\nu} = h^\mu_\mu =0,
\eeq
i.e.  the 5D ``gravity waves.''
The aim of our investigation can be summarized in the following question: \\

\vspace{-0.3cm}
{\flushleft {\em Are there asymptotically $AdS_5$ backgrounds   that support normalizable tensor zero-modes
($\Box_4 h_{\mu\nu} (x,y)=0$)  ?}} \\

The interest in requiring $AdS_5$ asymptotics in the UV is that, in this way, the 4D dual
theory is automatically UV-complete, and free of elementary gravitons, as we will
see in more detail in the next Section. Therefore, if a spin-2 zero mode   exists,
it must correspond to a composite state in a UV-complete non-gravitational theory.

\section{Spin-2 states in RS-like models}
\setcounter{equation}{0}

Let us examine this question in the simplest possible model: pure $AdS_5$ (better,
its Poincar\'e patch) with no boundaries, i.e. the metric given in (\ref{ads})
with  $y_{UV}=0$ and
$y_{IR}\to +\infty$. In this metric, the equation for the transverse traceless metric
fluctuations, defined by $\delta g_{\mu\nu} = (k y)^{-2} h_{\mu\nu}$, is \cite{RS1}:
\beq
h_{\mu\nu}'' - {3\over y} h_{\mu\nu}' + \Box_4   h_{\mu\nu} = 0.
\eeq
Zero mode solutions have the general  form
\beq
h_{\mu\nu}(x,y) =  h_{\mu\nu}^{(0)}(x) + y^4 h_{\mu\nu}^{(4)}(x),  \qquad \Box_4 h_{\mu\nu}^{(0)}(x) = \Box_4 h_{\mu\nu}^{(4)}(x) =0.
\eeq
In the $AdS$/CFT language, the perturbation $h^{(0)}_{\mu\nu}$ corresponds to turning on a
source for the stress tensor in the UV, whereas $h^{(4)}_{\mu\nu}$ corresponds to an IR deformation that gives
the stress tensor a VEV.

We can check normalizability of $h_{\mu\nu}^{(0)}$ and   $h_{\mu\nu}^{(4)}$ by expanding Einstein's action to second
order around the $AdS$ solution, and isolating  the kinetic terms for the tensor fluctuations:
\bea
&& S_{kin}[h^{(0)}]= \int_0^{+\infty} {dy\over (ky)^3} \int d^4 x\left(\de_\mu h^{(0)}_{\mu\nu}\right)^2, \\
&& S_{kin}[h^{(4)}]= \int_0^{+\infty} {dy \,y^4\over (ky)^3} \int d^4 x\left(\de_\mu h^{(4)}_{\mu\nu}\right)^2.
\eea
Clearly the $y$-integrals diverge for both $h^{(0)}_{\mu\nu}$ {\em and} $h^{(4)}_{\mu\nu}$: the former
is non-normalizable in the UV, the latter in the IR. This model does not contain
Spin-2 zero-modes.  The fact that $h^{(0)}_{\mu\nu}$ is not UV-normalizable is consistent
with the fact that it corresponds to an external non-dynamical source in the UV.

It is easy to  modify the above set-up in order to obtain a normalizable
tensor zero-mode: it is enough
to introduce a UV cut-off, and restrict the fifth dimension to the region $y \geq y_{UV} > 0$.
In this way we obtain the  the RS2 model, in which the mode  $h^{(0)}_{\mu\nu}$
is normalizable
and mediates four-dimensional gravity. In the dual description,  we have made the
external source  dynamical by adding to it a kinetic term in the fundamental UV
Lagrangian.

Thus, In the RS2 model  we did obtain a four-dimensional graviton, but in a trivial way:
 we have added the graviton as an extra fundamental degree of freedom to the
theory. A much more radical picture  would result, were we able to make the
mode $h^{(4)}_{\mu\nu}$ normalizable: since this is peaked in the IR, the
resulting state  would be interpreted as a composite graviton. We can make the question
asked at end of the previous section more precise, and ask: \\

\vspace{-0.4cm}
{\flushleft{\em Are there asymptotically $AdS_5$  backgrounds
 that support normalizable zero-modes
of the type $h_{\mu\nu}^{(4)}$ ?}}\\

To make  $h^{(4)}_{\mu\nu}$ normalizable,
 the UV cut-off is irrelevant (and we will always remove it
from now on), but we have to modify the theory in the IR. The simplest thing one can
do is to introduce a cut-off in the IR, and to impose appropriate boundary conditions
at $y_{IR}$ such that they are satisfied by $h^{(4)}_{\mu\nu}$. This is a particular case of
the more general models constructed by Gherghetta, Peloso and Poppitz \cite{gpp},
that consider a slice of $AdS_5$ bounded by both IR and UV boundaries, and introduce
Pauli-Fierz mass terms both on the branes and in the bulk to control which combination
of the graviton zero-modes is in the spectrum.

The approach taken in Ref. \cite{gpp} is not completely satisfying, as it is subject to the general
critique that applies to holographic models with a hard IR cut-off: the fifth
dimension terminates abruptly, and not by a dynamical process. In other
words, the strong coupling dynamics that leads to the IR scale is completely
disconnected from the UV (which in fact has unbroken conformal invariance).

Instead, we would like to consider more general models, in which deviations
from $AdS$ in the IR occur dynamically and can be obtained using the 5D field
equations.  This general analysis was performed in Ref. \cite{kn}, and
a review of the results found there  will be the subject of the rest of
the present work.

\section{Searching for  Spin-2  States in General 5D Backgrounds}
\setcounter{equation}{0}

We aim at investigating the existence of normalizable spin 2 zero-modes in the
most general five-dimensional, asymptotically $AdS$ geometry that respects four-dimensional Lorentz invariance.
In conformal coordinates~\footnote{I will use 5D coordinates $x^A = \{y,x^\mu\}$;
four-dimensional indices will always
be contracted with the flat metric $\eta_{\mu\nu}$;
 a prime will denote derivative w.r.t. $y$.} the metric reads:
\beq\label{metric}
ds^2 = a^2(y) \left(dy^2 + \eta_{\mu\nu}dx^\mu dx^\nu\right),   \qquad y>0
\eeq
and has  an $AdS_5$-like boundary with curvature scale $k$  at $y=0$,
\beq
 a(y) \sim {1\over ky}  \qquad  y\to 0.
\eeq

It is useful, although not strictly necessary if we  are only interested in tensor modes,
to work with a concrete five-dimensional setup.
If we assume the additional requirement that the space-time respects the Null Energy Condition (NEC)\footnote{This is the weakest of the energy conditions, and dropping it typically results
in pathological behavior such as violation of causality and  presence of ghosts.}, we can realize
any metric of the form (\ref{metric}) as a solution of an Einstein-Dilaton system with an
appropriate potential. Therefore, we take the 5D dynamics to be described by the action:
\beq \label{action}
S = {1\over 2\kappa_5^2 }\int d^5 x\, \sqrt{-g} \Big(R - g^{AB}\de_A \Phi \de_B \Phi - V(\Phi)\Big),
\eeq
The background solution is specified by the scale factor $a(y)$ and by the scalar field
profile $\Phi(y)$.
Choosing an appropriate potential $V(\Phi)$ one can obtain any
metric of the  form (\ref{metric}) that satisfies the NEC (see Refs. \cite{skenderis,freedman}).
$AdS$ asymptotics require that, as  $y\to 0$:
\beq
\Phi(y) \to \Phi_0, \quad V(\Phi_0) = 2\Lambda \equiv -12 k^2.
\eeq

Consider now  tensor fluctuations around the background metric, defined by:
\beq
ds^2_{pert} =  a^2(y) \left[dy^2 + \left(\eta_{\mu\nu} + h_{\mu\nu}(x,y)\right)dx^\mu dx^\nu\right], \quad h_{\mu}^{\mu} =  \de^\mu h_{\mu\nu}=0.
\eeq
 The  action (\ref{action}), expanded to second order in  $h_{\mu\nu}$
around the solution, reads:
\bea\label{quadaction}
S^{(2)}[h_{\mu\nu}] && = -{1\over 8\kappa_5^2}\int dy\, d^4 x\,
\sqrt{-g} g^{AB} \de_A h_{\mu\nu} \de_B  h^{\mu\nu} \nonumber \\
&& = -{1\over 8\kappa_5^2}\int dy\, d^4 x\, a^3(y) \left[\left(h_{\mu\nu}'\right)^2 +
\left(\de_\rho h_{\mu\nu}\right)^2\right].
\eea

The corresponding linearized field equation is:
\beq\label{spin-2}
h_{\mu\nu}'' + 3 {a'\over a} h_{\mu\nu}' + \Box_4  h_{\mu\nu} = 0.
\eeq
Notice that it is only the scale factor that enters this equation, not the
scalar field profile. Therefore, the properties of the tensor modes depend only
on the background metric, and not on the details of the underlying model of which it
is a solution. We can change the matter content by adding more fields, or replace
the scalar field by e.g. a perfect fluid, but all conclusions about these modes
will still hold. These details {\em do} matter for other modes, like scalar and
vector modes. I will briefly discuss these lower spin fluctuations in Section 6.

The problem of finding solutions  with a given 4D mass can be reduced to
a one-dimensional
quantum mechanical problem in the $y$-coordinate.  To see this,
assume  a factorized  ansatz:
\beq
h_{\mu\nu}(x,y) = h(y) h_{\mu\nu}^{(4d)}(x), \qquad \Box_4 h_{\mu\nu}^{(4d)}(x)
 = m^2  h_{\mu\nu}^{(4d)}(x),
\eeq
and define the ``wave-function'' $\Psi(y)$ as:
\beq
\Psi(y) = e^{-B(y)} h(y), \qquad B(y) \equiv -{3\over 2} \log a(y).
\eeq
In these variables, the tensor mode equation (\ref{spin-2}) becomes the Schr\"odinger-like
equation:
\beq\label{schr}
-\Psi'' + \left[ (B')^2 - B''\right]\Psi = m^2 \Psi.
\eeq

As in the previous section, the normalizability condition  can be read-off from the quadratic action (\ref{quadaction}):
\beq
 S \sim \int d y e^{-2B} h^2(y)  \int d^4 x  \left(\de_\rho h_{\mu\nu}^{(4d)}\right)^2  = \left(\int dy |\Psi|^2\right) \int d^4 x \left(\de_\rho h_{\mu\nu}^{(4d)}\right)^2.
\eeq
 Requiring finiteness of the effective four-dimensional kinetic term for $h_{\mu\nu}^{(4d)}(x)$, imposes
the standard normalizability condition on the wave-function $\Psi(y)$, i.e. it must be
square-integrable.

The problem of finding spin-2 zero modes is thus reduced to finding the zero-energy
eigenstates, ${\cal H}\Psi=0$,  of the Hamiltonian
\beq\label{hamil}
{\cal H} = -\de^2_y + {\cal V}(y), \qquad {\cal V}(y) \equiv  (B')^2 - B''.
\eeq
on the Hilbert space of square-integrable functions $\Psi(y)$. The Hamiltonian is completely
specified by the scale factor $a = e^{-2B/3 }$, and
we want to determine what are the  backgrounds such 
that zero-energy eigenstates exist.

While in general it is not possible to determine explicitly  the spectrum of ${\cal H}$, for
any choice of  $B(y)$ the two independent
zero-energy solutions of the differential equation  (\ref{schr})
can be written explicitly, and are given by:
\beq\label{zeromodes}
\Psi^{UV}(y) = e^{-B(y)}, \qquad   \Psi^{IR}(y) = e^{-B(y)}\int_0^y dy'\, e^{2B(y')}.
\eeq
The integration constant in the second solution has been chosen such that $\Psi^{IR}$ vanishes
at $y=0$: recall that $AdS$ asymptotics require $B(y)\sim -3/2\log y$ as $y\to 0$, therefore close
to the $AdS$ boundary:
\beq
 \Psi^{UV}(y) \sim y^{-3/2}, \qquad \Psi^{IR}(y) \sim y^{5/2}, \qquad y\to 0.
\eeq
Thus, $\Psi^{IR}$ is normalizable around $y=0$, whereas $\Psi^{UV}$ is not\footnote{Both
become normalizable if, as in RS, we introduce a cut-off and restrict $y> 1/\Lambda$.}.

Up to now, we have one candidate spin-2 zero mode, normalizable in the UV, whose
profile is given by $\Psi^{IR}$ in eq. (\ref{zeromodes}).  Whether or not this is
indeed an eigenstate of ${\cal H}$ depends on the behavior of $B(y)$ in the IR, i.e.
for large $y$.  From the holographic perspective, having zero-modes in the spectrum depends
on the infrared dynamics. Notice that, by definition,   $\Psi^{IR}$ vanishes in the UV,
so the corresponding state, if it exists, is composite rather than fundamental, since
it ceases to exist at high energy.

As a first step in the analysis, one can separate two classes of backgrounds:
\begin{enumerate}
\item the $y$ coordinate ranges from $0$ to $+\infty$;
\item space-time ends at a singularity or a boundary at a finite $y_0$.
\end{enumerate}

It is fairly easy to see that, in the first case, $\Psi^{IR}$ cannot possibly be normalizable
as $y\to \infty$: in fact it even fails to go to zero in that limit, no
matter what is the asymptotic behavior of $B(y)$. Thus, from now on we discard this possibility.

In the second  class of models, we certainly can have zero modes
if we truncate the space with a hard boundary ({\em hard wall}) at finite $y$. This is rather artificial, though,
and there are more general possibilities: for example the space-time  can terminate
dynamically by developing  a singularity at finite $y$. An IR singularity, unlike a hard boundary,
 always arises as the consequence of the running
of some coupling (in our case, the non-trivial background scalar field $\Phi(y)$),
and it seems to be inevitable in 5D holographic models with  a non-trivial
IR dynamics (for example, some recently proposed realistic-looking
5D holographic duals of confining gauge theories also have IR singularities \cite{gk1,gkn2}).

What kind of singular behavior of $B(y)$ results in a normalizable $\Psi^{IR}(y)$? Suppose,
for example,
that the singularity is due to the vanishing of the scale factor at $y=y_0$. Then, as $y\to y_0$,
$B(y)\to +\infty$, and
\beq
\Psi^{IR}(y) \sim (y_0-y) e^{B(y)}.
\eeq
Requiring $\Psi^{IR}$ to be normalizable around $y_0$,  we obtain the constraint:
\beq
\Psi^{IR} \prec {1\over(y_0-y)^{1/2}} \quad \Rightarrow \quad e^{B(y)} \prec {1\over(y_0-y)^{3/2}},
\eeq
i.e. $B(y)$ should diverge at most as fast as $-3/2 \log (y_0-y)$.

The other possibility is that $B(y)$ remains finite at $y_0$, and the singularity is due
to a divergence of one of its derivatives. In this case $\Psi^{IR}$ has a finite limit
and it is trivially normalizable.

Therefore, using the relation $a = e^{-2B/3}$, we can make the
following general statement: \\

{\em In a five-dimensional, asymptotically $AdS$ background, there
can be normalizable spin-2 zero-modes if the space-time ends
at some finite value $y_0$ of the conformal coordinate, where the scale factor $a(y)$
either stays finite, or vanishes more slowly  than $(y_0-y)$.}

\section{The Need for IR Boundary Conditions}
\setcounter{equation}{0}

The  space-times supporting spin-2 zero modes are either singular, or  terminated by a
hard wall in the IR. The presence of a singularity typically means that the
description being used breaks down, and one has to get a better understanding of
the infrared dynamics. One can hope that the singularity will be resolved in a full string
theory realization. However,  it is possible that the results obtained using the
singular theory  do not depend too much on the details of the dynamics that
regularizes the space-time. This is the case, for example, in models when the
singularity is ``screened'' from physical fluctuations: the wave functions do
not probe the high curvature region, and any modification in this region
will not drastically modify the spectrum of low-lying states  \cite{gkn2}.
 As I will  discuss
below, in the scenario discussed here we are not so lucky, and it is
not possible to completely decouple the singularity.

In gauge/gravity duality, one expects the properties of
the field theory to be completely defined by the gravity action, plus a set of boundary conditions
in the UV.
In the case of a  hard wall , we need to supply extra information  to specify both
 the wall position and the IR boundary conditions for the various fields (equivalently, 
 one has the freedom to add boundary terms  to the action in the IR). The UV data are not
  enough to completely determine the spectrum.  Since the normalizable wave-functions $\Psi^{IR}$ generically
 do not vanish in the IR, changing the details of the wall will affect the spectrum.

When the fifth dimension ends in a singularity, on the other hand, the position of
the singularity is completely determined by the UV asymptotics. In this case,
one can say that space-time ends dynamically. In certain cases,
the requirement of normalizability is sufficient to determine the spectrum
completely. If this is so,  the details of resolution of the singularity do not
significantly affect the spectrum\footnote{In the same  sense that,
for example,  the QCD details of the proton structure  
are  irrelevant for the  spectrum
of the hydrogen atom}.

 In other cases however, one
still has  to provide additional information in the IR, in the form of
``boundary conditions'' at the singularity, and this
is precisely what happens in the singular backgrounds that support normalizable
spin-2 zero-modes. Let us go back to the eigenvalue equation, eq. (\ref{schr}), and suppose
the singularity at $y=y_0$ is of the form:
\beq\label{alpha}
B(y) \sim -\alpha \log(y_0-y), \qquad a(y) \sim (y_0-y)^{2\alpha/3}, \qquad \alpha\geq 0
\eeq
As we have seen in Section 3, the  zero-modes $\Psi^{IR}$ are normalizable
if and only if $0<\alpha<3/2$\footnote{The case
$\alpha<0$ violates the NEC, and cannot be realized  in the model (\ref{action}), or in
general in any model without ghost-like degrees of freedom.}. For any $m^2$, close to the singularity
 eq (\ref{schr}) reads approximately:
\beq
 -\Psi'' - {\alpha^2 -\alpha \over (y_0 -y)^2}\Psi \sim 0,
\eeq
with the two independent solutions:
\beq\label{asymp}
\Psi \sim  c_1  (y_0 -y)^\alpha + c_2  (y_0 -y)^{1-\alpha},
\eeq
for arbitrary coefficients $c_1$ and $c_2$. For $0<\alpha<3/2$, i.e. when 
spin-2 zero-modes are allowed,
{\em both solutions are normalizable}. This means that normalizability
at the singularity does not impose any restriction
on the eigenfunctions, and a solution to the {\em full} equation exists
for {\em any} value of $m^2$ (in particular $m^2=0$). This does not mean that
the model has a continuous spectrum. Rather,  the spectral problem is
not fully specified by the information supplied so far.

From  a more formal
point of view,  the spectral problem is not fully specified, because
the Hamiltonian (\ref{hamil}) is not essentially self-adjoint. Rather, it
is a symmetric operator possessing infinite non-equivalent self-adjoint extensions,
each with a different spectrum. Picking one of these extensions is equivalent
to specifying the asymptotic behavior of the solutions  at $y_0$, i.e.
  by fixing the ratio of the coefficients $c_1$ and $c_2$  appearing
in (\ref{asymp}). There exists one specific choice for which the zero-mode is
in the spectrum, but generically this will not be the case.

The situation with the singularity, therefore, is not so different from the hard wall
scenario. In both cases there are normalizable, IR-localized  spin-2 zero modes only for
very special choices of the IR boundary conditions.

Notice that, in the singular case, when $1/2<\a<3/2$, or more
generally when 
$$-\log (y_0-y) \prec B(y) \prec 3/2 \log(y_0-y),$$ the
metric perturbation $h(y) = e^{B(y)} \Psi(y)$ diverges at the singularity, and
the linearized analysis cannot be trusted. On the other hand, when $\a<1/2$, the
metric perturbation reaches the finite value $c_1$ at the singularity,
so the linear approximation remains reliable.

\section{Lower Spin Modes}
\setcounter{equation}{0}

So far I have discussed the tensor  spectrum, to answer the
question whether  one can realize holographically a scenario where massless gravitons
are emergent. However, for the viability of the model, it is important to check
whether there also exist massless lower spin states. In particular, we want
to exclude models with scalar\footnote{Here, the terms
``scalar'' and ``vector'' refer to the transformation properties
under the 4D Lorentz group.} zero-modes (such as the radion in RS,
or the extra scalar mode in DGP ) that
couple with gravitational strength, since these are typically bad for phenomenology.
Massless transverse  vector modes, on the other hand, can be tolerated, since they
do not couple to a conserved stress tensor.
Here I will only sketch the main results, and refer the reader to Ref. \cite{kn} for
details.

\subsection{Scalars}

As we have seen, the spectrum of tensor
fluctuations is rather model-independent, and depends only on the background metric. On the
other hand
 the spectrum of  scalar fluctuations  generically depends on the full background.
 Moreover, of all the scalar fluctuation one can write, only a certain number
 corresponds to physical modes, the others can be gauged away using (linearized)
 diffeomorphisms (whereas tensor fluctuations are gauge-invariant).

 In a model with a single scalar coupled to gravity, such as (\ref{action}) one can show that,
 in the massive sector, there is a single physical scalar fluctuation, which solves a Schr\"odinger
 equation similar to eq. (\ref{schr}), but with $B\to B - \log[a \Phi'/a']$. In the
 massless sector things are different  due to  an enhanced gauge invariance, and as shown
 in Ref. \cite{kn} there are two independent fluctuations $\zeta_1, \zeta_2$, which solve the equations:
 \beq
 \zeta_1' = 0; \quad \left({a^4\over a'} \zeta_2 \right)' = -2 a^3 \zeta_1.
 \eeq
 The most general scalar zero mode, then, has the form:
\beq
 \zeta_1(x,y) = F(x); \qquad \zeta_2 = {a' \over a^4}(y)G(x) -2 \left({a'\over a^4}
 \int  a^3 \right)(y) F(x)
 \eeq
 where $F(x)$ and $G(x)$ are two arbitrary function satisfying $\Box_4 F,G=0$.

 Careful analysis of the effective kinetic term for $F(x)$ and $G(x)$ reveals that $F$ is
 not UV-normalizable, and $G$ (the would-be radion in the case of RS) is not IR-normalizable
  around  singularities that allow spin 2 zero modes. On the other hand, in hard
  wall models it must be killed by boundary conditions.

  \subsection{Vectors}

  In this sector there are only zero-modes: there do not exist transverse vector fluctuations
  at the massive level. The extra  scalar and vector that
  appear  in the massless sector become, in the massive sector,  the three
additional degrees of
  freedom of the  massive tensor  fluctuations (five d.o.f., as opposed to two for the
  massless tensor fluctuations). Therefore in each sector we have a total of six
  physical d.o.f., arranged in different representations of the four-dimensional
  Lorentz group.

  The  transverse vector equation is simply:
  \beq
  \left(a^3(y) A_\mu(x,y)\right) ' = 0  \quad \Rightarrow \quad A_\mu(x,y) = a^{-3}(y) v_\mu(x).
  \eeq
  The effective action for this mode is the 4D Maxwell action for $v_\mu(x)$, with the
  prefactor $\int dy  a^{-3}(y)$. This mode is UV-normalizable in asymptotically $AdS$
  backgrounds, and it is IR normalizable around singularities where that $a(y)$ vanishes
  more slowly than $(y_0-y)^{1/3}$. This overlaps, but does not coincide, with the behavior
  that allows spin-2 zero-modes. \\

  \subsection{Summary of Results}

  The results about
  the existence or non-existence of zero modes of various spins in IR-singular,
  UV-$AdS$ space-times is summarized in Table 1.
\begin{table}[h]
\begin{center}
\begin{tabular}{|c|c|c|c|c|c|c|}
\hline
\multicolumn{1}{|c|}{ }
&
\multicolumn{1}{|c|}{$y\in (0,\infty)$}
&
\multicolumn{5}{|c|}{$y\in (0,y_0)$}\\
\hline
\multicolumn{1}{|c|}{$B(y_0)$}&
\multicolumn{1}{|c|}{$-$}&
\multicolumn{3}{|c|}{$-\a\, log (y_0-y)$}&
\multicolumn{2}{|c|}{$finite + (y_0-y)^\b$}\\
& &$0<\a<1/2$ &$1/2<\a<1$&$1<\a<3/2$&$0<\b<1$&$1<\b<2$\\
\hline
Spin-2& $-$&$\bigcirc$&$\bigcirc$&$\bigcirc$&$\bigcirc$&$\bigcirc$\\
\hline
Spin-1& $-$&$\bigcirc$&$-$&$-$&$\bigcirc$&$\bigcirc$\\
\hline
Spin-0& $-$&$-$&$-$&$-$&$-$&$\bigcirc$\\
\hline
\end{tabular}
\end{center}
\caption{4D massless spectrum as a function of the IR behavior of $B(y)$.
A line $-$ or a circle $\bigcirc$ indicate  absence or presence, respectively,  of the corresponding
normalizable zero-mode. For the parameter ranges not shown in the table there is no normalizable 
spin-2 zero-mode.The last three lines indicate the behavior of the Shr\"odinger potential, the tensor mode wave-function and the scalar
 curvature near the end of the $y$-coordinate range.}
\end{table}

As remarked in Section 5, the case $1/2<\a<3/2$ is problematic, since close
to the singularity the metric perturbations diverge.

\section{Coupling Emergent Gravitons to 4D Matter}
\setcounter{equation}{0}

 In the setup 
described above, four-dimensional matter  can be coupled to gravity by adding
a probe 3-brane at a fixed position $y_b$, where the Standard Model is localized.
If we couple ordinary matter to the induced metric on the brane,
\beq
S_{matter} = \int d^4 x \sqrt{-g_{ind}} {\cal L_{SM}},
\eeq
the 5D action for linearized gravity plus localized matter will be:
\beq\label{matteraction}
S = -{1\over 8 \kappa_5^2}\int d y {a^3(y)\over a^2(y_b)} \left(\de_\rho h_{\mu\nu}(x,y)\right)^2 +
\int_{y=y_b} h_{\mu\nu}(y_b) T^{\mu\nu},
\eeq
where we have rescaled to  physical coordinates, $x^\mu \to a^{-2}(y_b)x^\mu$, so that
the background metric on the brane is $\eta_{\mu\nu}$.

The effective 4D gravitational coupling will  depend on the graviton
profile evaluated at  $y=y_b$. Inserting the definition,
\beq
h_{\mu\nu}(x,y) = a^{-3/2}(y)\Psi(y)h_{\mu\nu}^{(4d)}(x),
\eeq
we get from (\ref{matteraction}) the effective action for the four-dimensional  graviton $h_{\mu\nu}^{(4d)}(x)$ coupled
to matter:
\beq
S= -\left({1\over 8 k_5^2 a^2(y_b)} \int d y \, |\Psi|^2 \right) \int  d^4 x\,
\left(\de_\rho h_{\mu\nu}^{(4d)}\right)^2 + a^{-3/2}(y_b) \Psi(y_b) \int d^4 x\,h_{\mu\nu}^{(4d)}T^{\mu\nu}.
\eeq
The effective four-dimensional Newton's  constant can be read-off from the above equation (using
unit-norm wave-functions):
\beq
\sqrt{8\pi G_N} = k_5 \, {\Psi(y_b)\over \sqrt{a(y_b)}}
\eeq

The gravitational coupling  vanishes as $y_b^{3}$ as $y_b\to 0$, and placing
the SM brane close enough to the $AdS_5$ boundary, $y=0$, the four-dimensional
Planck scale can be made parametrically larger  than the typical Kaluza-Klein masses,
which is  generically of order $1/y_0^2$.

\section{Generalizations}
\setcounter{equation}{0}

We have seen that, using holographic constructions,  one does not  generically obtain IR-localized, normalizable
 massless spin-2 modes in four-dimensions. These modes exist only for a small class of IR singularities,
and only if special boundary conditions are imposed at the end of space. These singularities are not
of the ``good'' kind, since they are not screened from propagating modes, and
 the spectrum heavily depends on the details of the singularity.

Nevertheless, one can
investigate if there exist constructions that result in  a regularized version of these singularities, and
whether one can (more or less naturally) obtain
the right boundary conditions that yield massless
gravitons in the regularized models. One step in this direction could be the string theory constructions
described in Ref. \cite{sfetsos}, which start from continuous distributions of D3-branes in type IIB,
and upon reduction to 5D yield precisely a singularity of the right kind (e.g. one
can get a scale factor that behaves as in (\ref{alpha}), with $\a=1/2$.).

The difficulties we encountered are not surprising, in view of
 the Weinberg-Witten theorem \cite{ww}. This
states that, in four dimensions, one cannot have massless spin-2 modes that couple universally
to a Lorentz-covariant stress tensor\footnote{Ordinary gravity evades this theorem because the gravitational
stress tensor does not transform covariantly under a Lorentz transformation (it does only  up to a (linearized) diffeomorphism, which mixes its  
helicity 2 components with those of lower helicity).}. What we are looking
for after all,  via a holographic detour,
is composite massless  spin-2 states in a strongly coupled field
theory defined by its gravity dual, so we are running against the Weinberg-Witten theorem.

However, it is not obvious from the start that the Weinberg-Witten
theorem should apply to these holographic constructions, since the IR-localized  graviton zero modes, when  it exists,
will   not  couple  universally to the fundamental degrees of freedom. The source for the field
theory stress tensor is encoded in the non-normalizable mode in the UV, and this is the mode
that couples universally to the stress tensor at all energies. An example is  the zero-mode
in the Randall-Sundrum models, which has a constant profile, and the universality of its coupling
in 4D descends from the same property in 5D.
On the other hand, the UV-normalizable mode cannot interact  as a 4D graviton at all energy scales.
In fact, it becomes soft and  decouples from the fundamental theory at high  energy,
 (in particular the high energy behavior of these gravitons violates the equivalence principle).

Although the Weinberg-Witten theorem does not straightforwardly apply to our case, one
can guess that  that some of the difficulties we have encountered can be traced to that argument.
In order to alleviate these difficulties, therefore, one can try to lift some
of the hypotheses of that theorem, and at the same time make less stringent requirements
on the ``emergent'' spin-2 states we want to construct. There are (at least) two possible
roads one can follow:

\begin{enumerate}

\item Giving up masslessness, and  looking  for 5D space-times
with normalizable spin-2 states  that have a tiny 4D mass.
\item Giving up normalizability, and allowing the graviton to be a very long lived resonance.

\end{enumerate}

Interestingly, both ideas (massive and/or metastable gravitons) have received
a lot of attention from the phenomenological point of view, but at the same time  present
several unresolved theoretical challenges. The  holographic approach can probably
lead, for the first time, to manifestly consistent realizations of these kinds of  models.

\section*{Acknowledgments}

I would like to thank Elias Kiritsis for suggestions and comments.
The author is  supported by European Commission Marie Curie
Postdoctoral Fellowships, under contract number  MEIF-CT-2006-039369.
This work was also partially supported by
INTAS grant, 03-51-6346, RTN contracts MRTN-CT-2004-005104 and
MRTN-CT-2004-503369, CNRS PICS \#~2530,  3059 and 3747,
 and by a European Union Excellence Grant,
MEXT-CT-2003-509661.

\end{document}